\documentclass[twocolumn]{revtex4}

\usepackage{graphicx}

\def\a{\alpha}
\def\b{\beta}
\def\ga{\gamma}
\def\de{\delta}

\def\^#1{\widehat{#1}}

\def\beql#1{\begin{equation} \label{#1}}
\def\beq{\begin{equation}}
\def\eeq{\end{equation}}

\def\<{\langle}
\def\>{\rangle}

\def\({\left(}
\def\){\right)}

\def\[{\left[}
\def\]{\right]}

\def\eqref#1{(\ref{#1})}

\begin{document}

\title{Asymptomatic infectives and $R_0$ for COVID}

\author{Giuseppe Gaeta}

\address{Dipartimento di Matematica, Universit\`a degli Studi di Milano \\ via Saldini 50, 20133 Milano (Italy) \\ and \\ SMRI,  00058 Santa Marinella (Italy) \\ {\tt giuseppe.gaeta@unimi.it} }



\begin{abstract}
We discuss how the presence of a large set of asymptomatic infectives changes our estimate of the COVID-19 basic reproduction number, also known as $R_0$.
\end{abstract}

\maketitle

\section{SIR model and basic reproduction number}

The most basic tool for modeling epidemic behavior is the SIR model \cite{KMK,Murray,Edel,Heth,Britton}, partitioning the population into Susceptibles, Infected and Infectives, and Removed; in the COVID context, Removed means either healed (or dead) or isolated. The SIR model is -- in physicists' language -- a mean field one: all individuals are assumed to be equal and interact in the same way with any other one. These assumptions are of course not realistic, but the model ins an important tool to get some intuition about general mechanisms, also present in more refined ones \cite{Murray}.

The SIR equations are
\begin{eqnarray}
dS/dt &=& - \, \a \, S \, I \nonumber \\
dI/dt &=& \a \, S \, I \ - \ \b \, I \label{eq:SIR} \\
dR/dt &=& \b \, I \ . \nonumber \end{eqnarray}
The parameter $\b$ is the removal rate, and can be thought of as the inverse of the infection time (time from infection to removal); the parameter $\a$ takes into account many factors, such as the capacity of the virus to infect an organism  it gets in contact with, the individual protection measures, and the intensity of social contacts.

The number of infectives will raise as long as
\beql{eq:ga} S(t) \ > \ \frac{\b}{\a} \ := \ \ga \ ; \eeq this number $\ga$ is thus the \emph{epidemic threshold}.

In the SIR equations, the term $\a S I$ represents the new infected per unit of time; this means that each infective gives origin to
$$ \delta I \ = \ \a \ S \ (\de t) $$ new infections in the time $\de t$. As each infective is active, on the average, for a time $\b^{-1}$, and in the early phase we can take $S \approx S_0$, this means that each infective will give raise in this phase to
\beql{eq:R0} R_0 \ = \ \frac{\a}{\b} \ S_0 \ = \ \frac{S_0}{\ga} \eeq new infections. This number is the \emph{basic reproduction number} for the model. (The notation $R_0$ is maybe unfortunate, as it may seem to refer to the initial datum for $R(t)$, but it is traditional and we will keep to it; moreover in the case of COVID -- as far as we know -- there is no natural immunity, so $R(0)=0$ and no confusion can arise.) In the case of COVID-19, estimates of $R_0$ from epidemiological data suggest $R_0 \simeq 2.5 - 3$; this can be compared with $R_0$ for standard seasonal flu, which is about half.

\section{Asymptomatic infectives}

It is by now clear that in the case of COVID there is a large set of \emph{asymptomatic infectives}. We want to discuss how this affects our estimate of $R_0$.

In a recent contribution \cite{Gasir} I have introduced a modified version of the SIR model, taking into account the relevant presence of asymptomatic infectives and thus called A-SIR model. In this, there are two classes of infected/infectives individuals, $I$ and $J$, and two classes of removed ones, $R$ and $U$. Here $I$ represents the known infectives, $J$ the unknown (in particular, asymptomatic) ones; similarly $R$ represents the registered recovered individuals, while $U$ the unregistered ones -- basically those who went through an asymptomatic infection and are removed from the epidemic dynamics only once they are naturally healed. The model assumes that both classes of infectives are equally infective (it would be easy to formulate a variation removing this assumption, but we want to deal with the simplest model accounting for asymptomatic infectives); on the other hand, while symptomatic infectives are promptly removed from the dynamics by Hospital or home isolation, asymptomatic ones stay around for all the infective period. Thus the A-SIR equations are
\begin{eqnarray}
dS/dt &=& - \, \a \ S \, (I +J) \nonumber \\
dI/dt &=& \a \, \xi \ S \, (I+J) \ - \ \b \, I \nonumber \\
dJ/dt &=& \a \, (1-\xi) \ S \,(I+J) \ - \ \eta \, J \label{eq:ASIR} \\
dR/dt &=& \b \, I \nonumber \\
dU/dt &=& \eta \, J \ . \nonumber \end{eqnarray}
Note that the last two equations (like the last one for SIR) amount to direct integrations,
$R(t) = R_0 + \b \int_{t_0}^t I(y) d y$, $U(t) = U_0  + \eta \int_{t_0}^t J(y) d y. $

Here the parameter $\b$ represents again  the inverse of the removal time for registered infectives, while $\eta$ represents the removal time for unregistered infectives. In practice, $\b^{-1}$ corresponds to incubation time (first COVID symptoms appear usually after about 5 days) plus some delay for these to be recognized as such; our fitting of early data for the epidemics in Northern Italy gave the value $\b^{-1} \simeq 7$ days. On the other hand, $\eta^{-1}$ represents the removal time for undetected infectives; this corresponds to the incubation time plus the time needed for the organism to spontaneously cancel the infection, and our (clinically reasonable) working hypothesis in \cite{Gasir} was $\eta^{-1} \simeq 21$ days.

Here the number of new infected per unit of time is $\a S (I+J)$, and again in the early phase of the epidemic we can assume $S \approx S_0$. Thus each infective will give origin in the time span $\de t$ to
$\a S_0 \de t $ new infectives; we assume that each of these will be registered -- and thus isolated after an average time $\b^{-1}$ -- with probability $\xi$, while it will remain undetected -- and thus disappear from the epidemic dynamic -- with a probability $1-\xi$. Current estimates of $\xi$ range from $\xi = 1/10$ to $\xi = 1/7$ \cite{Li}, albeit smaller values have also been suggested \cite{Oxf}. It should be mentioned that the values given here proved to fit rather well (accompanied by a fit on the effect of restrictive measures) the data for Italy within the A-SIR framework; see \cite{Gasir}.

Thus we should look at the \emph{average removal rate} $B$ or equivalently to the \emph{average infective time} $B^{-1}$ in the early phase of the epidemic; in there the ratio between registered and total infectives is simply
\beq x \ := \ \frac{I}{I+J} \ = \ \xi \ , \eeq
while in later stages the proportion between $I$ and $J$ changes, as individuals stay longer in the $J$ class than in the $I$ class.

The average removal rate is
\beql{eq:B} B \ = \ \xi \, \b \ + \ (1-\xi) \, \eta \ . \eeq

This means that each (symptomatic or asymptomatic) infective individual will give direct origin, across its infective and non-isolation period, not to $R_0 = \a S_0 / \b$ but instead to
\beql{eq:R0h} \^R_0 \ = \ \frac{\a}{B} \ S_0 \ = \ \frac{\b}{B} \ R_0 \eeq
new infectives. As $\b > B$, this means that the actual basic reproduction number $\^R_0$ is larger --and possibly substantially larger -- than the value which is estimated solely on the basis of registered infections.

A trivial computation on the basis of the values given above -- i.e. $\b^{-1} \simeq 7$, $\eta^{-1} \simeq 21$, $\xi \simeq 1/10$ -- provides
\beql{eq:R0corr} \^R_0 \ = \ \frac{5}{2} \ R_0 \ . \eeq

This could explain why all Health Systems were surprised by the rapid growth of the number of COVID-19 infections; in fact, the presence of a large set of asymptomatic infectives was not realized when the epidemic attacked the first countries, and is becoming clearly established only now \cite{Li}, also thanks to the large scale epidemiological studies recently conducted in Italy \cite{Cri}.

\section{Estimating $R_0$}

\begin{figure}
    \includegraphics[width=180pt]{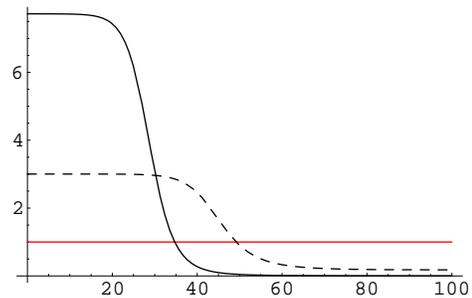}\\
  \caption{Time evolution of $\rho(t)$ with compatible initial conditions as predicted by the A-SIR (solid curve) and the standard SIR (dashed curve) models; the epidemic threshold level $\rho = 1$ is also shown (red). The numerical computations are carried out for $N=S_0 = 6*10^7$, $\a = 7.15*10^{-9}$, $\b = 1/7$, $\eta = 1/21$. Initial data are $I_0 = 100$ for both runs, with $J_0=900$ (corresponding to $\xi=1/10$) for the A-SIR run.}\label{fig:R0}
\end{figure}

It should be noted that this divergence between the SIR and the A-SIR reproduction number will also affect \emph{apriori} estimations of time evolution for the reproduction number $\rho(t)$, as illustrated in Fig.\ref{fig:R0}. In that case, the predictions about the time at which $\rho$ gets below $\rho=1$, and hence the epidemic begins extinguishing in a natural way, would differ (for data given in the caption to Fig.\ref{fig:R0} and inspired by the Italian epidemic) by about two weeks for the two models; this may be a very relevant difference for political decisions.

Note that in the early phase of the epidemic using standard SIR theory leads to an underestimation of $\rho$; while in later stages it leads to an overestimation of $\rho$. This is easily understood in qualitative terms: the underestimating in early phases is due to not considering the large number of asymptomatic infectives; on the other hand, at later stages this same large numbers contributes substantially to the depletion of the susceptibles reservoir, i.e. to taking the population below the epidemic threshold.

On the other hand, if estimates are not \emph{apriori} but are made on the basis of available data, the question is more delicate; we will now discuss it.

First of all we note that for a given level of susceptibles $S$, we always have a higher $\rho$ in  the A-SIR than in the SIR setting, i.e.
$$ \rho_{{\mathrm asir}} (S) \ = \ \frac{\a \, S \, (I+J)}{I \b \ + \ J \, \eta}  \ > \ \frac{\a \, S}{\b} \ = \ \rho_{{\mathrm sir}} (S) \ ; $$ this follows at once from elementary manipulations and by $\b > \eta$.

It is however improbable that decisions are made on the basis of $S(t)$, for the simple reason that this quantity cannot be directly measured. The only number which is known while the epidemic is running its course is the number of \emph{registered} infected; as these are promptly isolated, what we know is indeed $R(t)$. Thus we are called to evaluate $\rho (t)$ on the basis of our knowledge of $R(t)$.

Working in the framework of the SIR model, one would proceed as follows. The equations for $S$ and for $R$ yield $dS/dR = - S/\ga$ and hence
\beq S(R) \ = \ S_0 \ \exp[ - (R - R_0 )/\ga ] \ . \eeq
In the case of COVID, as no natural immunity exists, at the beginning of the epidemic we have $S_0= N$ (the whole population) and $R_0 = 0$; this yields the simpler formula
\beq S(R) \ = \ N \ \exp [ - R/\ga ] \ . \eeq
As the reproduction number $\rho$ is given by $\rho= S/\ga$ (see the discussion above), in practice this is obtained by the epidemiological data for $R$ as
\beql{eq:rhosir} \rho (t) \ = \ (N/\ga) \ \exp [-R(t) / \ga ] \ = \ \rho_0 \ e^{- R/\ga} \ . \eeq

We are not able to provide a similar closed formula for the A-SIR framework; we can however compare the result of numerical integration for the A-SIR model with what Health organizations would deduce from the \emph{same} data for the $R(t)$ time series using the SIR framework, i.e. eq.\eqref{eq:rhosir} above.

\begin{figure}
  \includegraphics[width=180pt]{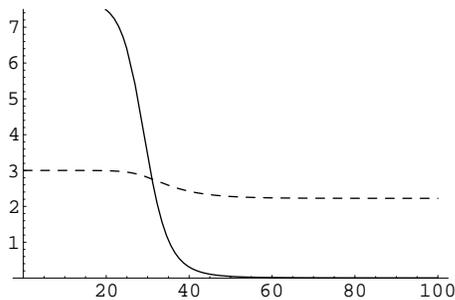}\\
  \caption{Estimating $\rho$ from a time series for $R(t)$.We have integrated the A-SIR equations with parameter values and initial conditions as in Fig.\ref{fig:R0}, and computed $\rho(t)$ from the numerical solution (solid curve); we have also computed the estimate of $\rho$ which would result by applying formula \eqref{eq:rhosir}, i.e. standard SIR theory (dashed).}\label{fig:est}
\end{figure}

It is clear from Fig.\ref{fig:est} that interpreting the data resulting from an epidemic evolution with a large class of asymptomatic infectives by standard SIR theory, i.e. disregarding asymptomatic infectives, not only leads to a gross error in estimating the reproduction number as the epidemic runs its course, but leads to considering that any spark could restart the epidemic in a situation where instead the population of susceptibles is well below the epidemic threshold, precisely due to all the people who got infected and recovered without being registered in the statistics.

With the parameters used in the numerical simulation of Fig.\ref{fig:est}, at the final stage we get $$ R/N \ \simeq \ 0.0999 \ , \ \ \ U/N \ \simeq \ 0.8995 \ . $$ Thus susceptibles represent only a very tiny fraction of the population, while an analysis based only on registered infections would estimate them to be about 90\% of the population. It is clear that this would lead to gross errors in making choices about very relevant aspects, such as maintaining restrictive measures.

Note also that the same mechanism seen earlier on, leading to underestimation of $\rho$ in early phases and overestimation of it at later stages, is present here.

\section{A concrete example: COVID in Italy}

\begin{figure}
  \includegraphics[width=180pt]{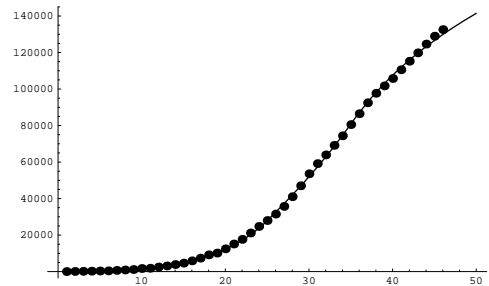}\\
  \caption{A-SIR model versus experimental data for COVID in Italy. The points represent cumulative $R(t)$ official data, while the line the best fit provided by the A-SIR model, see text. Reproduced from \cite{Gasir}.}\label{fig:ITA}
\end{figure}

In a recent work \cite{Gasir} we have shown that the A-SIR model describes rather well the development of the COVID epidemic in Italy, provided the contact rate $\a$ is changed at certain dates to take into account the enforcement of restrictive measures; see Fig.\ref{fig:ITA} for a plot of how the model fits experimental data.

In particular, in this case the model parameters are those used in Fig.\ref{fig:est}, except that now $\a = \a(t) = \mu (t) \a_0$, with $\a_0 = 7.15*10^{-9}$, and the modulation (piecewise constant) function $\mu(t)$ is given by
$$ \mu(t) \ = \ \cases{1 & for $t < t_1 \ = \ 25 $ \cr 0.5 & for $t_1 < t < t_2 = 35$ \cr 0.15 & for $t > t_2 $. \cr} $$
Here $t$ is measured in days, with $t=0$ at February 20, and $t_1$, $t_2$ correspond to one week (i.e. $\b^{-1}$ days) after the introduction of the two sets of restrictive measures by the Italian government on March 8 and March 22; these values were obtained by fitting experimental data. See \cite{Gasir} for details.

\begin{figure}
  \includegraphics[width=180pt]{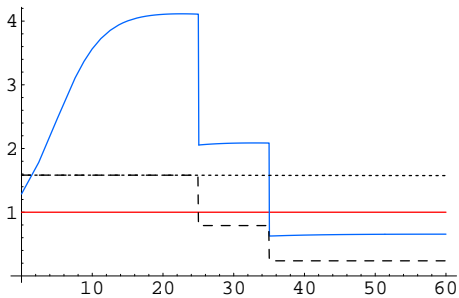}\\
  \caption{Different estimations of $\rho (t)$ from the time series of $R(t)$ for the COVID-19 epidemic in Italy. Red: the level $\rho = 1$ discriminating between self-sustained and self-extinguishing epidemic; Blue: the ``true'' $\rho(t)$, computed from numerical solution of the A-SIR equations fitting experimental data (see Fig.\ref{fig:ITA}); Black, dotted: ``naive'' estimation of $\rho$ from the SIR model from eq.\eqref{eq:rhosir} with constant $\ga$; Black, dashed: estimation from eq.\eqref{eq:rhosir} now taking into account the change of $\ga = \b / \a(t)$. See text.} \label{fig:comp}
\end{figure}

We can now apply the same analysis as above on our model using these concrete data. The outcome of this is illustrated in Fig.\ref{fig:comp}.

It should be noted that this figure points out different dangers related to estimating $\rho$ from epidemiological data and an interpretation of these based on the SIR model. In fact, for $t_1 < t < t_2$ an estimate based on considering the achieved reduction of $\a$ would lead to consider the situation as a safe one when it is not so according to the A-SIR results; on the other hand, based on SIR but not assuming $\a$ has changed would still lead to the correct conclusion that the epidemic is expanding. On the other hand, for $t>t_2$ the situation is safe according to the A-SIR model, and also according to the SIR model taking into account the changes in $\a(t)$; but it would be considered as not safe based on the ``naive'' SIR model, i.e. not taking into account the changes in $\a(t)$.

Needless to say, decisions based on a wrong estimate of $\rho$ can be dangerous (in different ways) in both cases, either allowing a new spark of infection or forcing an unneeded long lockdown.

\section{Conclusions}

Much attention is devoted, also in popular media and when it comes to political decisions about measures designed to counter epidemic spreading, to the basic reproduction number -- often denoted as $R_0$ and which we called $\rho (t)$ in our discussion. Albeit this is strictly speaking a constant, to be evaluated at the very beginning of the epidemic, it  is understood that this parameter will change -- both due to restrictive measures and by the natural epidemic dynamic -- over time; fundamental decisions can depend on our estimate of the level reached by $R_0$, in particular depending if this is above or under the critical value $R_0 = 1$, which discriminates between epidemic spreading or the infection extinction.

Motivated by the peculiar features of the COVID-19 ongoing epidemics, and in particular by the presence of a \emph{large class of asymptomatic infectives}, we have analyzed how taking this characteristic into account modifies our estimates of $R_0$, both at the initial stages of the epidemic and over its development.

We have shown -- considering for the sake of concreteness the parameters which apply to the COVID epidemic in Italy, as determined in a recent paper of ours \cite{Gasir} -- that this can require a correction to the initial $R_0$ as large as $5/2$, see eq.\eqref{eq:R0corr}. This may explain why most Nations were caught short by the very fast COVID expansion in the early stages of the epidemic.

We have then analyzed how disregarding the presence of this large set of asymptomatic infectives can also lead to mis-estimate the reproduction number at later stages of the epidemic. In particular we have compared the \emph{apriori} predictions and the \emph{aposteriori} estimates based on field data which arise from the A-SIR model \cite{Gasir} and thus taking into account this peculiar feature of the COVID epidemic with those which would arise from a straightforward application of standard SIR theory. We found that even in this context lack of appreciation of the role of asymptomatic infectives would lead to gross errors in estimating the reproduction number and thus possibly also in decision making on how to contrast the epidemic.

\bigskip
\section{Acknowledgements}
The work was carried out in lockdown at SMRI. I am also a member of GNFM-INdAM.

\end{document}